\documentstyle[6175,times]{l-aa}
\input{psfig}

\begin{document}

\title{Stellar populations from adaptive optics observations: four test
cases}

\author{
	T. R. Bedding	\inst{1,2} \and
	D. Minniti	\inst{2,3}	\and
	F. Courbin	\inst{2,4,5}	\and
	B. Sams \inst{6,7}
}

\institute{
Chatterton Astronomy Department, School of Physics, University of Sydney
2006, Australia 
\and
European Southern Observatory, Karl-Schwarzschild-Str.~2, D-85748 Garching
bei M\"{u}nchen, Germany
\and
IGPP, Lawrence Livermore National Laboratory, P.O. Box 808, MS L-413, Livermore, CA 94550
\and
Institut d'Astrophysique, 5 Avenue de Cointe, B-4000 Liege, Belgium
\and
URA 173 CNRS-DAEC, Observatoire de Paris, F-92195 Meudon Principal
C\'edex, France
\and
Max Plank Institut f\"ur Astrophysik, Karl-Schwarzschild-Str.~1, D-85748
Garching bei M\"{u}nchen, Germany
\and
Current address: mediateam, Weidenweg 2c, 85375 Neufahrn, Germany
}

\date{Received Feb 17, 1997; Accepted June 5, 1997}

\offprints{\\Tim~Bedding~({\tt bedding@physics.usyd.edu.au})}

\maketitle

\markboth{T.R. Bedding et al.: Stellar populations from adaptive optics}{}

\begin{abstract} 

We describe a first attempt to apply adaptive optics to the study of
resolved stellar populations in galaxies.  Advantages over traditional
approaches are (i)~improved spatial resolution and point-source sensitivity
through adaptive optics, and (ii)~use of the near-infrared region, where
the peak of the spectral energy distribution for old populations is found.
Disadvantages are the small area covered and the need for excellent seeing.
We made observations with the ADONIS system at the European Southern
Observatory of the peculiar elliptical galaxy NGC~5128; the irregular
galaxy IC~5152 (a possible outer member of the Local Group); the Sc galaxy
NGC~300 (a member of the Sculptor group); and the Sgr window in the bulge
of the Milky Way.  These different fields give excellent test cases for the
potential of adaptive optics.  In the first two cases, we failed to obtain
photometry of individual stars, which would have required excellent seeing.
For NGC~300 we measured magnitudes for nine individual supergiants ($H =
18.3$--20.2), but did not go deep enough to detect the tip of the RGB of an
old population.  For the Sgr field we produced a infrared luminosity
function and colour-magnitude diagram for 70 stars down to $K\simeq 19.5$.
These are the deepest yet measured for the Galactic bulge, reaching beyond
the turn-off.

\keywords{Instrumentation: Adaptive Optics -- 
Galaxies: stellar content --
Galaxies: individual: NGC~5128, IC~5152, NGC~300  --
Galaxy: stellar content
}
\end{abstract}

\section{Introduction}

Much progress has been made towards an understanding of galaxy formation
\citeeg{B+D96}.  However, the formation of elliptical galaxies and the
bulges of spirals and the relationship between them remains an area of
considerable uncertainty.  Did they form shortly after the Big Bang in a
short period of intense activity that consumed all the gas and prevented
further star formation?  Or did they have several episodes of star
formation, somehow triggered by mergers or accretion processes?  In this
case, the observed light would come from both old and intermediate age
stars.

The issue can be resolved by constructing luminosity functions (LFs) and
colour-magnitude diagrams (CMDs) for the stars in ellipticals and bulges,
and comparing these with predictions from stellar evolutionary theory.  The
theory is fairly well understood, having been tested and calibrated on open
and globular clusters spanning a wide range of ages and metallicities (see
reviews by \citebare{R+FP88}; \citebare{VdBSB96}).  The project requires
photometry to identify and measure luminosities for the stars at the top of
the asymptotic giant branch (AGB).  This method would complement the
existing integrated spectroscopic data and would allow one to study the
stellar distribution, to separate the different populations, and to
estimate the ages, metallicities and distances of these populations.

HST/WFPC2 has been used for such projects at visible wavelengths, both for
globular clusters in nearby galaxies (\citebare{FPBC96}; Jablonka et al.,
in preparation) and the stellar components of more distant galaxies
\cite{Fre92,SMW96}.  However, there are several arguments for making these
stellar population studies in the near-infrared \cite{Sil96,M+B95}:
\begin{itemize}

\item The spectral energy distributions of red giant stars peak in the
infrared, increasing the contrast relative to the underlying fainter and
bluer stars.

\item Extinction and reddening from dust are less than in the visible ($A_K
= 0.1 A_V$).

\item The degeneracy in the optical colours of the red giant branch is
avoided.  This degeneracy makes it difficult to determine ages or
metallicities from optical photometry, especially for the more metal-rich
populations that dominate bulges and ellipticals \cite{BBO91}.

\item The transformation between the photometric observations and theory
($M_{\rm bol}$, $T_{\rm eff}$) is easier in the infrared than in the
visible: the $J-K$ colour is directly related to $T_{\rm eff}$, and the $H$
and $K$ magnitudes are directly related to $M_{\rm bol}$ (e.g. Frogel et
al. 1992, Johnson 1992).  The $H$ filter is particularly useful for
constructing deep luminosity functions because the bolometric correction is
essentially independent of colour ($BC_H = 0.26 \pm 0.10$;
\citebare{B+W84}).

\end{itemize}

Until now, accurate infrared photometry for individual stars has only been
possible in galaxies of the Local Group.  The bulge of the Milky Way has
been well studied in the near-infrared \cite{F+Wh87,Dav91,MOR95}, as have
the dwarf elliptical M32 \cite{Fre92} and the bulges of the Local Group
spirals M31 \cite{RMG93} and M33 \cite{MOR93,McL+L96}.  Obtaining CMDs for
more distant galaxies would clearly be a very significant result.  For
example, the top of the AGB in M31, M32 and M33 is clearly seen to have a
sharp cutoff at $K\simeq16$ \cite{M+B95}, reflecting the fact that these
three galaxies are at about the same distance.  Measuring this cutoff
magnitude in more distant galaxies should allow distance determination.
Here we make steps towards this goal by exploiting another advantage of
near-infrared observations: the availability of adaptive optics.

Adaptive optics (AO) involves using a deformable mirror to make real-time
corrections to image distortions that arise from atmospheric turbulence
(see \citebare{Bec93} for a recent review).  By reducing the diameters of
stellar images and so increasing the signal relative to the background sky,
this technique gives a substantial gain in the sensitivity of point-source
photometry.  Furthermore, the improvement in spatial resolution reduces
confusion in crowded regions.  One important limitation is that, until laser
beacons become available, we are restricted to fields which lie near
a bright foreground star.

In this paper we describe a first attempt to apply adaptive optics to the
study of resolved stellar populations in galaxies.  We have selected four
widely different targets for this study which together provide excellent
cases for testing the potential of adaptive optics.

\section{Observations and data reduction}

Observations were made in 1995 March and August with the ESO 3.6-m
telescope using ADONIS (ADaptive Optics Near Infrared System;
\citebare{RBH94}; \citebare{BBC94}; \citebare{G+L95}).  The ADONIS
instrument is the successor to COMEON+.  Wavefront sensing is done at
visible wavelengths and the science detector operates in the near
infrared\@.  For the latter we used the SHARP~II camera, which contains a
$256\times 256$ NICMOS~3 array with $0\farcs05$ pixels, giving a field of
$12\farcs5 \times 12\farcs5$.  The response of the system peaks in the $H$ band
(1.6\,$\mu$m).

An adjustable mirror in ADONIS allows one to make offsets of a few
arcseconds within a field, which we found useful in constructing a local
flat field.  It also allowed us to ensure that neither the bright reference
star nor its diffraction spikes fell on the science detector.  For each
field we obtained a sequence of exposures offset by $2''$ from the
field centre in each of four orthogonal directions.  The median of these
frames was used to construct a flat field.  After dark-subtraction,
flat-fielding and interpolation over hot pixels, the images were aligned
and added to produce a single mosaic.  Due to the sub-stepping process, the
central $8''\times8''$ of the mosaic has the greatest effective exposure
time and hence the highest signal-to-noise ratio.

We performed aperture photometry using the DAOPHOT package within IRAF\@.
Our photometric standard was HD~161743 (spectral type B9\,IV), for which we
adopted magnitudes of $K = 7.62$ and $H-K = 0.00$ \cite{EFM82}.  The
internal errors were measured using DAOPHOT in all cases.  Results for the
four targets are described in the next section.

\section{Results}

\subsection{The giant elliptical galaxy NGC~5128}

The S0/E~pec.\ galaxy NGC~5128 (Cen~A) is of interest both as the closest
radio galaxy and because it shows evidence of having undergone a recent
interaction.  Fortuitously, two foreground stars are superimposed on the
body of the galaxy.  These are SAO~224118 ($V=8.5$, spectral type K5) and
SAO~224131 ($V=9.1$; F0), which lie at distances of 4.5$'$ and 10.9$'$ from
the nucleus and away from the central dust lane.

The ADONIS observations were made on 1995 March 23 and August 21.  We
observed fields next to both reference stars, but seeing conditions were
poor and fast variations prevented good correction.  The first star,
SAO~224118, was bright ($K \simeq 5$) and produced too much scattered light
on the science detector, preventing accurate sky subtraction.  {}From a
total integration time of 1~hour near the second star we failed to detect
any individual stars in NGC~5128, down to an estimated magnitude limit of
$K = 19.5$.  Assuming the presence of old giant stars reaching $M_K = -5$,
this allows us to place a lower limit on the distance to the galaxy of
about 3\,Mpc.

Since these observations, \citeone{SMW96} have published the first
colour-magnitude diagram of the NGC 5128 halo, obtained with HST/WFPC2\@.
They resolved this galaxy into stars and convincingly detected the tip of
the old RGB at $I = 24.1$, deriving a distance of $3.6\pm0.2$\,Mpc,
consistent with our lower limit.  These RGB stars would have $K \approx
22.5$, beyond our magnitude limit.  \citename{SMW96} also detected a
handful of AGB stars, extending to $I = 22.6$, equivalent to $K \approx
20.5$.  These stars would be at the limit of our detection and their
absence is explained by the small field covered here, although we speculate
that a few would have been detected with better seeing.

\subsection{The halo of IC~5152}

IC~5152 is a typical dwarf irregular (dIrr) galaxy on the outer fringe of
the Local Group, at a distance of about 1.6~Mpc \cite{San86,vdBer94}.  We
observed a field in the outer part of IC~5152 in 1995 August using SAO
247284 as a reference.  This star has coordinates $\alpha_{2000} = 22^{\rm
h}\,02^{\rm m}\,36\fm5$, $\delta_{2000} = -51^{\circ}\,16'56\farcs5$,
magnitudes $V=7.9$, $B-V=0.1$, $K\simeq 7$ and spectral type~A3\@.  The
seeing was 0\farcs6--1\farcs0.  {}From a 40-minute integration at $H$ we
detected three very faint stellar sources but low counts prevented us from
deriving accurate magnitudes.

Our failure to detect more stars could indicate that IC~5152 is more
distant than 1.6\,Mpc, or that its halo does not extend this far from the
nucleus.  On the basis of these marginal detections and the results
obtained on NGC~300 (discussed below), we believe that useful results would
have been possible on IC~5152 under excellent seeing conditions.

\subsection{The disk of NGC~300}	\label{sec.NGC300}

NGC~300 is a spiral galaxy in the Sculptor Group and lies at a distance of
about 2\,Mpc \cite{FMH92,Wal95}.  Infrared photometry of the brightest
M-supergiants in NGC~300 has been carried out by \citeone{H+G86}.  They
published colours and magnitudes for ten stars, five of which they
confirmed spectroscopically as galaxy members, with the rest being
foreground dwarfs.  The magnitudes of these bright M-supergiants in NGC~300
range from $K = 14.81$ to $15.51$ and their colours range from $H-K = 0.10$
to $0.28$.  

There are no suitable reference stars for the bulge of NGC~300.  We used a
bright star superimposed on the disk, about 2$'$ SW of the nucleus in the
prominent SW spiral arm.  The star is located at $\alpha_{2000} = 00^{\rm
h}\,54^{\rm m}\,46\fs4$, $\delta_{2000} = -37^{\circ}\,43'\,07\farcs0$ and,
according to the Guide Star Catalogue, has a magnitude of 9.2 on a IIIaJ
plate.  In the future, with improved AO systems, it may be possible to use
the compact nucleus of NGC~300 itself as a reference to study the central
regions.

We used ADONIS on 1995 August 24 to observe a field 15$''$S and 15$''$E of
the reference star.  For these observations, the offsets applied between
each sub-exposure were 1$''$ rather than $2''$.  The field was located in
between, but did not include, the red stars R21 ($V = 20.4$) and R20 ($V =
18.9$) of \citeone{H+G86}.

\begin{figure}
\centerline{
\psfig{figure=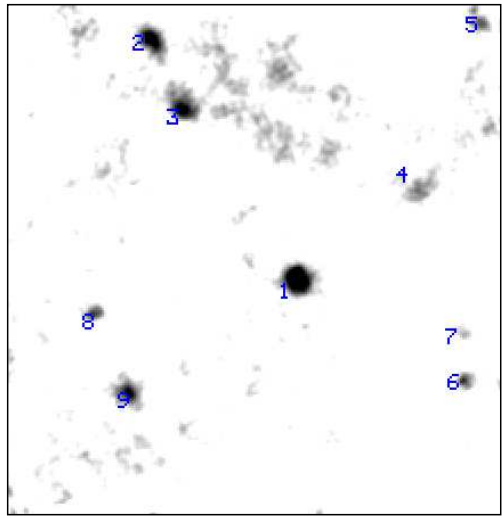,bbllx=89pt,bblly=172pt,bburx=520pt,bbury=621pt
,width=65mm}}
\caption[]{\label{fig.NGC300-ADONIS} ADONIS $H$-band image of the NGC~300
disk field.  North is up and east is to the left.  The image is $8''\times
8''$.  }
\end{figure}

A $K'$ image taken during particularly good seeing ($0\farcs6$ at the
seeing monitor) revealed about twelves sources with FWHM
$0\farcs3$--$0\farcs4$.  However, reliable photometry was prevented by a
bright strip across the image that was probably caused by reflected or
scattered light from the reference star.  For comparison, a deep
(30-minute) NTT frame taken in $I$~band in 0\farcs6 seeing as part of a
separate program \cite{ZMB96} shows the same stars, confirming the
sensitivity of adaptive optics.  Note that photometry of the NTT image in
this region is impossible because the reference star is saturated.

Our $H$-band ADONIS image was not compromised by scattered light and is
shown in Figure~\ref{fig.NGC300-ADONIS}.  It is the result of a 40-minute
exposure taken in moderately good seeing (1$''$ at the seeing monitor).
However, the slow variation of the seeing allowed good correction by the AO
system.  The $H$ magnitudes of the stars labelled 1 to 9 are: 18.29, 18.54,
19.16, 20.21, 19.41, 19.48, 19.76, 19.81 and 19.67.  The stars detected
here are more than three magnitudes fainter than the bright supergiants of
\citeone{H+G86} that reach $M_K = -11$.

This field is not crowded, although the surface magnitude of the underlying
stars is $\mu_B = 22.1$\,mag/arcsec$^2$ \cite{Car85}.  We resolved
individual supergiants, but did not go deep enough to detect the tip of the
RGB of an old population, which would be at $H = 20.5$ at the distance of
NGC~300.  The stars observed  here correspond to the bright supergiants with
$I \approx 20.5$ seen in the optical colour-magnitude diagrams of the disk
of NGC~300 \cite{RPC85,ZMB96}.

\subsection{The faintest stars in the Galactic bulge}

We selected a field in the direction of the Galactic bulge, which should
yield information about the structure of the inner Milky Way.  We observed
the Sgr field, which is located at $(l,b) = (0\degr,-3\degr)$, at a
projected distance of $0.4$ kpc from the Galactic centre (assuming $R_0 =
8$ kpc).  This is a crowded field, with sources covering a wide range of
magnitudes.  The reference star is located at $\alpha_{2000} = 17^{\rm
h}\,58^{\rm m}\,52\fs0$, $\delta_{2000} = -29^\circ\,52'\,44''$.  This star
has estimated magnitudes of $K = 8$ and $V = 9.5$.

\begin{figure}
\centerline{
\psfig{figure=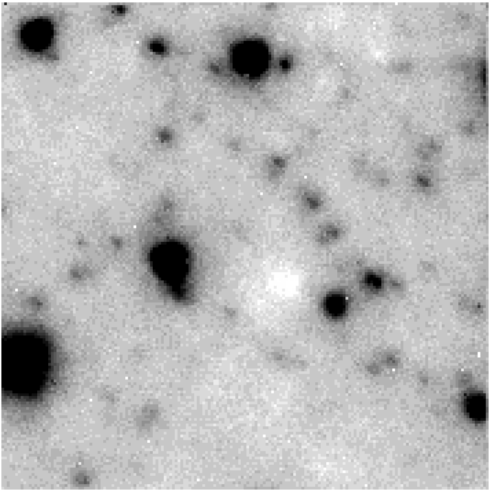,width=65mm}}
\caption[]{\label{fig.SGR-Kimage}
ADONIS $K'$ image of the Sgr window.  The image is $8''\times 8''$.}

\end{figure}

\begin{figure}
\centerline{
\psfig{figure=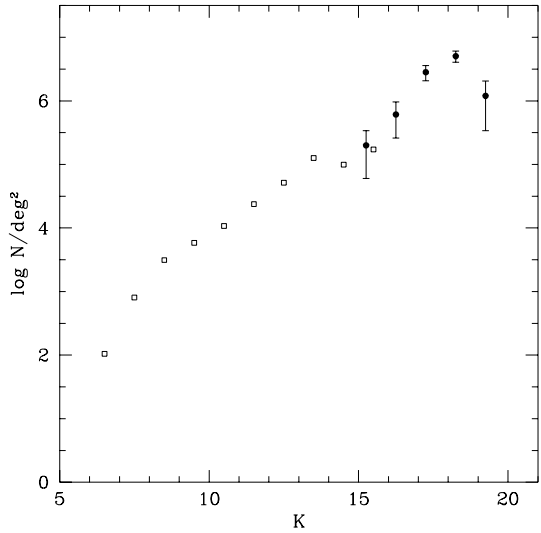,bbllx=30pt,bblly=159pt,bburx=562pt,bbury=687pt%
,width=70mm}}
\caption[]{\label{fig.SGR-LF} $K$ luminosity function for the Sgr window.
The squares are from \citeone{Min95} and the filled circles are from the
ADONIS observations.  The lower counts in the faintest bin indicate
incompleteness.  }

\end{figure}

Figure~\ref{fig.SGR-Kimage} shows a 40-minute exposure in $K'$ taken on
1995 August 24 during particularly good seeing.  The FWHM of stellar
sources is $0\farcs35$.  This image is much deeper than a combined exposure
of 160 minutes taken the previous night in poorer seeing.  We have found 70
stars in this image and obtained aperture photometry with DAOPHOT\@.
Because of the variation of the PSF across the image, the aperture
corrections for both the program stars and the photometric standard are
large ($\sim 0.6$ mag) and are therefore an important source of error in
the final photometry.

Note that the accuracies of photometry and of transformation to the
standard system are limited for AO observations by the fact that the AO
corrections differ for the program stars, the AO reference star and the
photometric standards, due to sky and seeing variations.  All images were
taking in a close temporal sequence to minimize these errors, but we
conclude that our zero point is good to only $\sim 0.2$ mag.

Figure~\ref{fig.SGR-LF} shows a $K$ luminosity function for the Sgr
window.  As well as the ADONIS data, we have included the measurements of
\citeone{Min95}, which cover the brighter magnitudes from $K = 6$ to $16$.
This infrared luminosity function represents the deepest yet measured for
the Galactic bulge, reaching beyond the bulge turn-off, which is located at
$K \approx 18$.

\begin{figure}
\centerline{
\psfig{figure=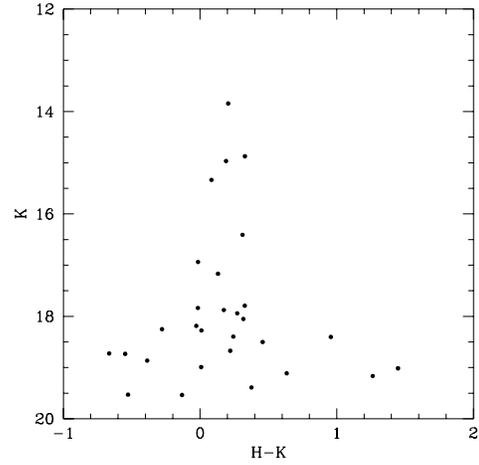,width=65mm}}
\caption[]{\label{fig.SGR-CMD} Infrared colour-magnitude diagram for the
Sgr window.  The main sequence turn-off of the bulge is at $K\approx 18$,
and the brightest stars in this figure lie on the sub-giant branch. }
\end{figure}

Figure~\ref{fig.SGR-CMD} shows a deep infrared colour-magnitude diagram
from our ADONIS data.  Note, however, that the $H$-band photometry is less
accurate because of poorer seeing and we detected only half of the sources
seen in the $K'$ images taken on August 22.  The position of the sources
for the photometry were determined in the deeper $K$ images.  The internal
errors in the $K$ magnitudes for $K<16$ are small (a few hundredths of a
magnitude), and for $K>18$ are much larger ($>0.5$ mag).  The $H$
photometry has larger errors, and the limiting $H$ magnitude is about one
magnitude brighter than at~$K$.

\section{Conclusions}

We have made a first attempt to apply adaptive optics to the study of
stellar populations in our galaxy and beyond.  In the cases of NGC~5128 and
IC~5152, we failed to detect individual stars.  We believe that these
targets should be feasible with ADONIS in good seeing.  For NGC~300 we
resolved a small number of K supergiants in the disk and presented a deep
$H$-band luminosity function.  For the Sgr bulge window our
colour-magnitude diagram and luminosity function are the deepest yet
obtained, reaching the turn-off of the bulge population for the first time
in the infrared.  These results demonstrate the feasibility of the method.
Four factors are important in determining potential results:
\begin{description}

\item[Seeing] The quality of adaptive correction depends critically on the
temporal frequency of the seeing variations.  Obtaining CMDs of distant
galaxies is only feasible under excellent seeing conditions ($\leq
0\farcs8$ and slow variations).  Good seeing, and hence good adaptive
correction, also helps to reduce the problems of field crowding.

\item[Contamination] The line of sight to distant objects goes through the
halo of our Galaxy, which contains many low mass stars.  These can be
estimated down to very faint magnitudes using simple Galactic models, and
turn out not to be important.  The contamination by background galaxies is
expected to dominate at very faint $K$ magnitudes.  In principle, this
could be accounted for by observing control fields.

\item[Availability of reference stars] We require a bright star to be
conveniently located for wavefront sensing.  We found it best to use
reference stars of $V\simeq10$.  Brighter stars produce stray light
contamination, making it difficult to flat-field and background-subtract,
and fainter stars do not allow good correction.

\item[Field of view] Adaptive optics correction is presently restricted to
a small field around the reference star.  This makes it time consuming to
cover large regions of sky, which is necessary to improve the statistics.

\end{description}

With ADONIS on the ESO 3.6-m telescope we reached a 3-sigma limiting
magnitude of $H = 20.0$ in one hour on point sources.  Based on the
limiting magnitude obtained in the NGC~300 field under good seeing
conditions (FWHM$\leq 0\farcs8$), we estimate that ADONIS can detect the
brightest stars in bulges and ellipticals out to a distance modulus of $m-M
\approx 27.5$ ($\sim$3\,Mpc).  These figures are based on an absolute
magnitude of $M_H = -8$ for the brightest giants in spheroidal populations
of Local Group galaxies (neglecting reddening).  For comparison, for the
Centaurus group has $m-M = 27.8$ \cite{SMW96}.

The results presented here should be surpassed by HST/NICMOS, provided that
this system can be made to operate to specification.  However, observations
using adaptive optics on 8-m class telescopes, while restricted to small
fields of view, should be competitive with HST/NICMOS because of the larger
aperture and higher spatial resolution.

Laser beacons should dramatically increase the sky coverage of adaptive
optics (see the review by \citebare{Bec93}).  A natural guide star will
still be required to determine the overall wavefront tilt, but this star
can be faint and the chances will be high of finding a suitable guide star
somewhere in the field of interest.  Using a laser beacon will also
eliminate the problem of scattered light from the bright reference star, an
important advantage.

We intend in the future to use the new deconvolution/co-addition codes
developed recently by \citeone{MCS97}.  Their algorithm optimally combines
numerous dithered frames of the same object in a deep sharpened image, on
which accurate photometry can be performed \citeeg{C+C97}.  It is very
likely that such a technique, combined to the improved performances of the
AO systems of the VLT, will be competitive with HST/NICMOS\@.

\begin{acknowledgements}

We are happy to thank the ADONIS team and the Adaptive Optics Group at ESO
for their efficiency and support, especially J.-L. Beuzit, P. Bouchet,
N. Hubin, D. Le Mignant, P. Prado and E. Prieto.  We also thank
A. Quirrenbach, A. Zijlstra and R. Fosbury for their help and useful
discussions.  T.R.B. is grateful to the Australian Research Council for
financial support.  F.C. is supported by ARC94/99-178 "Action de Recherche
Concert\'ee de la Communaut\'e Fran\c{c}aise (Belgium)" and P\^ole
d'Attraction Interuniversitaire P4/05 (SSTC, Belgium).  This work was
performed in part under the auspices of the U.S. Department of Energy by
Lawrence Livermore National Laboratory under Contract W-7405-Eng-48.

\end{acknowledgements}

\end{document}